# Spin Texture Control and Magnetic Gap Engineering in a Ferromagnetic Insulator-Topological Insulator Sandwiched Heterostructure.

*Mohammad T. H. Bhuiyan [1,2]†, Qile Li [1,3]†, James Blyth [1,4], Mengting Zhao [1,4,7], Ji-Eun Lee [5], Jonathan Denlinger [5], Jaime Sánchez-Barriga [6], Alexander Fedorov [6], Anton Tadich [7], Emile Rienks [6], Sung-Kwan Mo [5], Alexei Fedorov [5], Oliver J. Clark [1]*, Mark T. Edmonds [1,4]**

AUTHOR ADDRESS

[1] School of Physics and Astronomy, Monash University, Clayton, VIC 3800, Australia.

[2] Department of Physics, Pabna University of Science and Technology, Pabna 6600, Bangladesh.

[3] Department of Chemistry, Stanford University, CA 94305, United States.

[4] ARC Centre for Future Low Energy Electronics Technologies, Monash University, Clayton, VIC 3800, Australia.

[5] Advanced Light Source, LBNL, 1 Cyclotron Road, Berkeley, CA 94720, United States.

[6] BESSY-II, Helmholtz Zentrum Berlin, Albert-Einstein-Straße 15, 12489 Berlin, Germany.

[7] Australian Synchrotron, ANSTO, 800 Blackburn Rd, Clayton VIC 3168, Australia.

* Email: oliver.clark1@monash.edu

* Email: mark.edmonds@monash.edu

ABSTRACT - Quantum materials that combine magnetism with topological order are emerging as key platforms for next-generation spintronics and low-energy electronics. They enable the realization of emergent quantum phenomena such as the quantum anomalous Hall effect and axion insulator states. The ferromagnetic insulator (FMI)/topological insulator (TI)/FMI sandwich structure of single-septuple layer (1SL) $MnBi_2Te_4$/four-quintuple layer (4QL) $Bi_2Te_3$/1SL $MnBi_2Te_4$ holds great potential to achieve such desirable quantum phenomena at elevated temperature, owing to its large Dirac point band gap and high Curie temperature. Here, spin- and angle-resolved photoemission spectroscopy (spin-ARPES) is employed to directly verify that the band gap arises from broken time-reversal symmetry via proximity-driven magnetization. This study demonstrates direct control of the spin state via external magnetic fields and unambiguously confirms the exchange interaction as the gap-opening mechanism. The robust magnetic gap and controllable spin texture make this heterostructure a suitable candidate for spintronic applications and magnetic topological quantum phases.

The interplay between topology and magnetism in quantum materials gives rise to novel quantum phases, characterized by topologically protected surface states (SS) with non-trivial electronic band structures and complex spin textures.[1,2] One of the most compelling outcomes of this interplay is the quantum anomalous Hall effect (QAHE), where a single chiral edge mode enables dissipationless electron transport along the system edges without any external magnetic field. The realization of QAHE requires a magnetic exchange gap to open at the Dirac point (DP) of the surface states of a topological insulator (TI).[1,3-5] For example, $Bi_2Te_3$ is characterized by strong spin-orbit driven band inversion, leading to spin-momentum–locked, gapless Dirac SS protected by TRS, resulting in a chiral spin texture[1,3,6] as depicted in the left panel of Fig. 1a. When TRS is broken by coupling a topological insulator to a magnetic material, the exchange interaction can open a gap at the DP, producing a 1D chiral edge mode within the gap[4,7-11] depicted in the right panel of Fig. 1a.

A sandwich heterostructure composed of a four-quintuple layer (4QL) $Bi_2Te_3$ between two single septuple layers (1SL) of $MnBi_2Te_4$ (referred to as MBT/4BT/MBT for the remainder of this manuscript), shown in Figs. 1b and 1c, has recently been proposed as an magnetic topological insulator (MTI) through proximity induced TRS breaking.[12,13] Indeed, a large DP gap of 75 meV and relatively high Curie temperatures (20–30 K) have been reported,[12] making this MTI material a strong candidate for the realization of the QAHE at elevated temperatures. Despite the prior characterisation of the gap,[12] the spin texture of the exchange-split surface state remains an open, but crucial, question. In addition, several reports exist regarding the opening of gaps in $Bi_2Se_3$-family TIs via non-magnetic means,[14-16] and therefore the direct characterization of the MTI spin texture - and the ability to manipulate it with an external magnetic field - would provide an unambiguous origin for the observed gap.

In this study, we report a detailed characterization of the electronic band structure and surface-state spin texture of the MBT/4BT/MBT heterostructure using spin- and angle-resolved photoemission spectroscopy (spin-ARPES). Our findings reveal a clear spin splitting at the Γ-point and demonstrate direct manipulation of spin polarization using external magnetic fields, providing an unambiguous verification that the gap arises from exchange splitting rather than non-magnetic origins. In addition to this, the presence of chiral spin texture away from the Γ-point underscores the coexistence of magnetic and topological features in this MTI material.

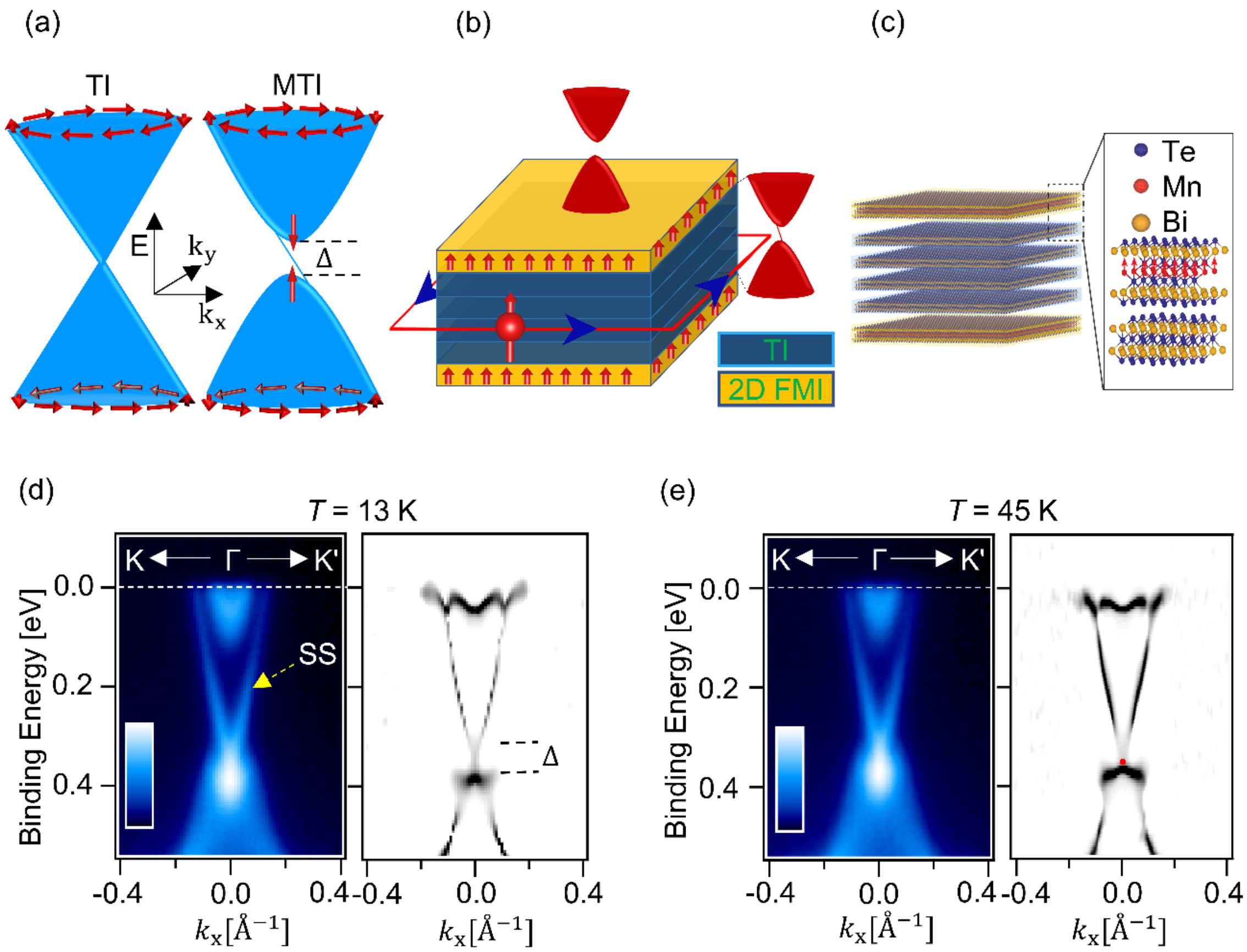

**Figure 1.** Interplay between topology and magnetism (a) Schematic band dispersion and spin textures of the SS in TIs and MTIs. (b) The magnetic exchange gap and spin-momentum locked chiral edge mode. (c) Structure of the thin film heterostructure: four quintuple layers of $Bi_2Te_3$ between two single-septuple layers of $MnBi_2Te_4$. Yellow sheets represent the ferromagnetic layers ($MnBi_2Te_4$), and blue sheets represent the TI layers ($Bi_2Te_3$). Red arrows (inset) show the magnetic moments of $Mn^{2+}$ in the $MnBi_2Te_4$ layer. (d,e) ARPES spectra and corresponding 2D curvature analysis[17] results taken at $T = 13$ K and 45 K, respectively, along the KΓK' direction at $h\nu = 47$ eV. The 2D plots in the right panels of d) and e) demonstrate the SS conduction and valence bands and corresponding gap opening (Δ = 72±12 meV) below Tc and how it closes above Tc (red dot indicates the gapless Dirac point).

High-quality MBT/4BT/MBT heterostructures were grown by molecular beam epitaxy (MBE) on Si(111) 7×7 substrates, following established methods.[12,18] High crystallinity and uniform sample surfaces were ensured by *in situ* reflection high-energy electron diffraction (RHEED) (see Fig. S1 in supporting information). To confirm the DP gap in this MTI material, a high-resolution ARPES spectrum was taken along the KΓK' high symmetry direction at 13 K (just below the Curie temperature) using 47 eV photon energy, as shown in Fig. 1d. The corresponding 2D curvature plot[17] in Fig. 1d confirms a large DP gap of 72±12 meV, consistent with the literature values[12]. Notably, this large bandgap diminishes above the Curie temperature, as shown in Fig. 1e which shows ARPES and corresponding 2D curvature plot at $T$ = 45 K, providing strong evidence that the gap originates from magnetism-induced effects. Further details regarding the gap analysis methodology are provided in the Supporting Information.

To probe the nature of the observed mass gap and to characterize the full surface state spin texture, we employ spin-ARPES to investigate the origin of the gap opening at the Dirac point (DP) in the MBT/4BT/MBT heterostructure, as shown in Fig. 2. In order to obtain a complete picture of the spin texture of the surface states of a proximity-induced magnetic topological insulator, we performed three-dimensional spin-resolved energy dispersion curves (spin EDCs) at Γ and at four further k-points symmetric about Γ, as indicated in Figs. 2a and 2d with dashed lines. Asymmetry in the up and down spin profiles is obtained by, $A = (I_{up}-I_{down})/(I_{up}+I_{down})$, and the spin polarization (P), shown in Figs. 2b)-c) and 2e)-f), is obtained by, P = A/S, where S = 0.25 is the Sherman function.

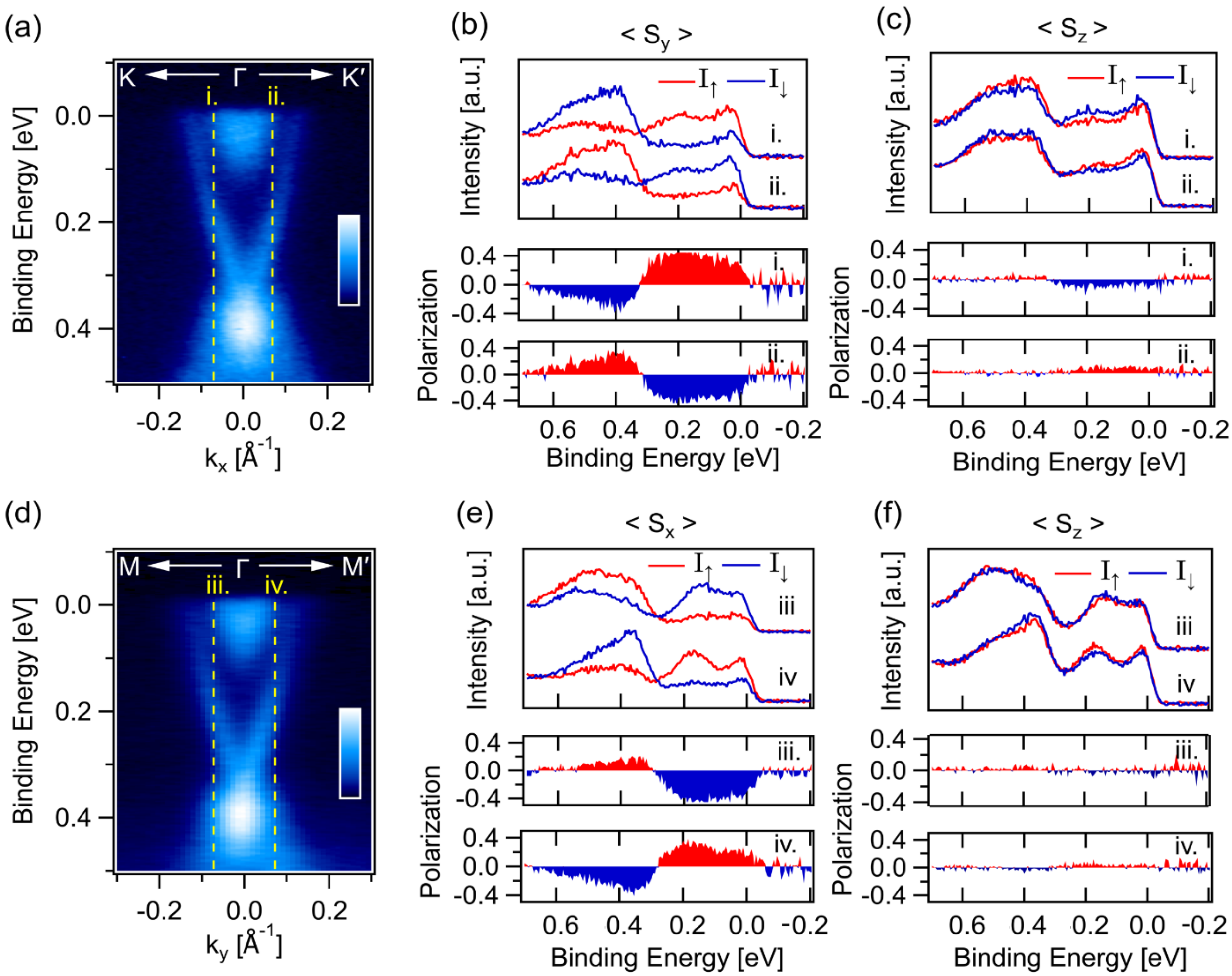

**Figure 2.** Spin texture of the MBT/4BT/MBT heterostructure away from the Γ-point. (a), (d) ARPES spectra along the KΓK' and MΓM' directions near the Brillouin zone center and Fermi surface. Yellow dashed lines indicate measurement positions for spin-resolved energy dispersion curves (EDC) shown in (b, c) and (e, f), corresponding to points (i), (ii), (iii), and (iv). (b), (c) Spin-resolved EDC and spin polarization (in-plane and out-of-plane) for positions (i) and (ii). (e), (f) Spin-resolved EDC and spin polarization (in-plane and out-of-plane) for positions (iii) and (iv). Measurements were taken at $T = 13$ K and $hv = 47$ eV.

Lorentzian fits to the pairs of spin-EDCs at points (i) through (iv) demonstrate a predominantly chiral spin texture: Along KΓK' (//$k_x$ direction), the y component of spin has polarization magnitudes of 71±13% and 65±14% above and below the DP. Along MΓM' (//$k_y$), the x component of spin becomes dominant with polarization magnitudes of 56±17% and 59±18% above and below the Dirac point respectively. This spin-momentum locked spin texture of the surface state is consistent with its assignment as a topological surface state,[3,19,20] though we note a small spin polarization along the out-of-plane direction is also evident in Figs. 2c and 2f. These signals also switch sign about $k_{\|}=0$, and thus arise from TRS-preserving mechanisms. This is consistent with the deviation to a fully chiral spin texture imposed by hexagonal warping of the constant-energy contours, as demonstrated in numerous topological systems.[21-23] In contrast, spin signals of magnetic origin are most likely to appear at the Γ-point where the exchange gap is present.[2,24] To probe this, we pulse-magnetize our samples with a ±0.5 T out-of-plane magnetic field (see methods) both to align magnetic domains[25] during spin-ARPES measurements under remanent magnetization, as well to enable control of magnetic spin signals through the reversal of applied magnetic pulse polarity. This approach thus enables the decoupling of magnetic and non-magnetic spin signals. In Figure 3, out-of-plane spin profiles and corresponding polarizations are shown for the Γ-point, following the application of oppositely oriented out-of-plane field pulses of ~0.5 T, measured with 27 eV photons and at 6 K (well below the Curie temperature), as illustrated in Figs. 3b and 3c. In Fig. 3(b, c), the thick solid lines are overlaid as guides to the eye extracted from fits to the experimental data to highlight the pronounced and robust spin polarization regions. Despite both the much lower efficiency of this mini-Mott system relative to the VLEED systems used for the data shown in Fig. 2, and the requirement for limited time-acquisitions in order to maintain the remanent magnetization (see Methods for discussion), Figs. 3b and 3c clearly show significant spin asymmetry and polarization at the gapped DP region, with the out-of-plane spin polarization flipping at the Γ-point as it crosses the Dirac gap (around 0.23 eV binding energy). Moreover, Fig.

3c illustrates that the out-of-plane spin polarization reverses direction with the opposite polarity magnetization, confirming its magnetic origin. The presence of spin switching about the Dirac point is consistent only with the presence of a magnetic exchange gap, where spin-polarization at k = 0 is otherwise strictly enforced to be zero by TRS. These findings are consistent with the magnetic gap and spin-splitting observed in thicker films of Mn-doped $Bi_2Te_3$.[26] We note that finite spin polarization was observed also in the bulk conduction bands, consistent with previous reports on similar systems.[27,28] Furthermore, to be consistent with these findings, an out-of-plane spin-EDC taken at an elevated temperature of 45 K (well above the sample's Curie temperature), shown in Fig. S3 in the supporting information, demonstrates an absence of finite spin polarization, consistent with the lack of ferromagnetic order above $T_C$ in the sample. These findings show that the band gap opening in MBT/4BT/MBT is magnetic in origin as a result of broken TRS. To summarize the full spin texture from Figs. 2 and 3, Fig. 3d shows the iso-energy contour of our magnetic TI over the SS region. The red dashed lines represent the Dirac-like electron and hole bands, while the red and yellow arrows indicate the chiral in-plane spin texture, highlighting the opposite spin polarization of the electron-like and hole-like bands. Green arrows at the Dirac gap crossing show the reversible spin orientation of the Dirac-like fermions at the Γ-point.

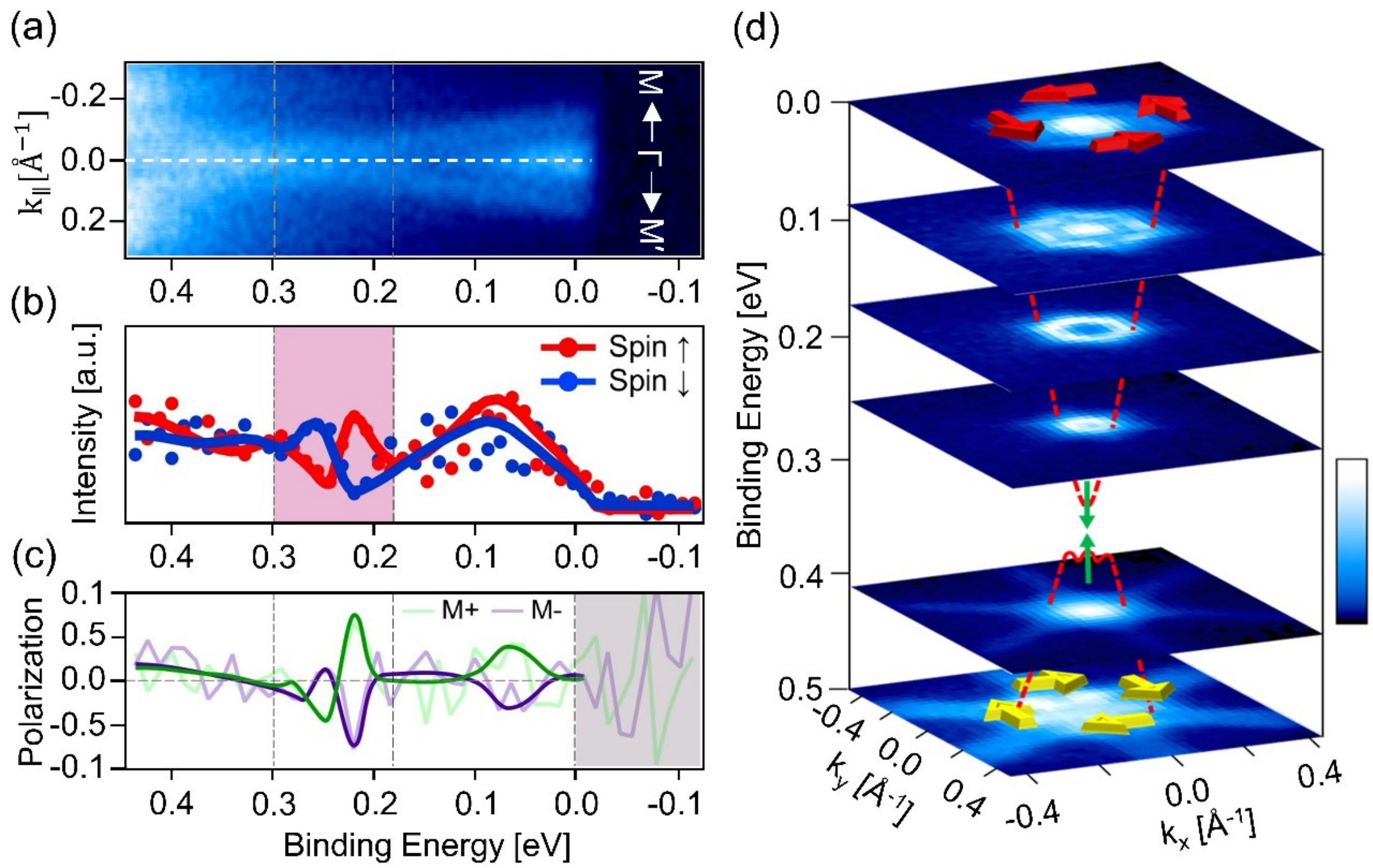

**Figure 3.** Confirming the magnetic origin of the gap. (a) Band dispersion of the heterostructure after magnetization. The white dashed line indicates the spin measuring point in momentum space. (b) Γ-point spin profiles of the out-of-plane spin component. (c) Spin polarization at the Γ-point: The green and violet curves represent the spin polarization corresponding to two opposite magnetization polarities. Thick solid lines in (b) and (c) are guides to the eye extracted from fits to the experimental data. The vertical dashed lines and pink-colored shaded area mark the gap region where spin polarization flipped. (d) Iso-energy contour of the heterostructure across the gap region with the spin-ARPES results. Red (yellow) arrows indicate the in-plane spin chirality of electrons (holes) away from the Γ-point, and the green arrows indicate the out-of-plane spin polarization at the gapped DP.

We have demonstrated clear control of the magnetic spin state in an MBT/4BT/MBT heterostructure, to provide unambiguous evidence that the 72 ± 12 meV gap originates from magnetic interactions, whilst also preserving the hallmark chiral spin texture of non-magnetic topological insulators away from the Γ-point. These findings provide important insights into the magnetic proximity effect in topological insulators and the delicate interplay between magnetism and non-trivial band topology. Furthermore, the ability to control the spin state with a 0.5 T pulse magnetic field makes magnetic-topological insulator heterostructures promising candidates for novel low-power spintronic device architectures and also paves the way for the advancement of lossless transport in topological insulators at higher temperatures.

EXPERIMENTAL METHODS

**Thin Film Heterostructure Growth:** High-quality thin films of MBT/4BT/MBT were grown using molecular beam epitaxy (MBE) on *n*-type Si(111) 7x7 substrates at a growth temperature of 230 °C under ultrahigh vacuum (UHV) conditions following an established recipe.[12,18] High purity Mn (99.9%), Bi (99.999%), and Te (99.95%) source materials were used in the effusion cells to deposit them on the substrate during growth. Deposition rates were calibrated by a quartz crystal microbalance (QCM) prior to the growth process. Layer-by-layer epitaxial growth of the heterostructure was monitored and controlled via in-situ reflection high-energy electron diffraction (RHEED). For beamline measurements at the Advanced Light Source (ALS), samples were grown in the MBE side chamber at the 10.0.1.1 endstation of ALS, with the pressure maintained below $8x10^{-10}$ mbar. For beamline measurements at BESSY-II Synchrotron, samples were grown using an Omicron Lab-10 MBE system at Monash University, with the pressure kept below $3x10^{-9}$ mbar. After growth, a 20 nm Tellurium capping layer was deposited to protect the films from contamination during transport to the spin-ARPES endstations at ALS and BESSY-II. After introduction into UHV, the tellurium capping layer was removed by mild-annealing prior to spin-ARPES measurements. The sample was gradually heated to a maximum temperature of 300 °C and held at this temperature for one hour to ensure the complete removal of the Te layer.

**Electronic Structure and Spin-resolved Measurements:** Spin-integrated electronic structure and spin-resolved measurements were carried out with *p*-polarized synchrotron light using the spin-ARPES on beamline 10.0.1.2 of the Advanced Light Source (ALS) at Lawrence Berkeley National Laboratory and beamline U125-PGM of the BESSY II Light Source at Helmholtz Zentrum Berlin. At ALS, data was taken using a Scienta-Omicron DA-30 spectrometer with two Ferrum VLEED detectors at 13 K (i.e., just below the magnetic ordering temperature). Measurements were carried out with two different photon energies (42 eV and 47 eV) to optimize the photoelectron signal. The spin-resolved energy dispersion curve (EDC) data was recorded along all three axes without moving the sample. Each spin-resolved EDC measurement was performed as a function of binding energy at a fixed momentum (k) with an energy resolution of 20 meV. The Sherman function value for the Ferrum VLEED detectors is 0.25. The out-of-plane spin-ARPES experiments at low temperature were conducted at the BESSY II Light Source (Helmholtz-Zentrum Berlin), using a dedicated in-vacuum pulse magnetization setup. In this setup, ±0.5 T magnetic field was applied *in situ* using a compact electromagnet at the measurement temperature. The sample was then moved down into the analysis position under remanent magnetization conditions for spin-ARPES measurements. This procedure was repeated for opposite out-of-plane magnetization directions. Data was taken using a Scienta R4000 spectrometer with two orthogonal mini-Mott detectors. To ensure the stability of the remanent magnetization over the course of the measurements, we carefully monitored the evolution of the spin polarization signals during extended acquisition sessions. The spin polarization remained robust and stable within the first 2 hours after magnetization, but began to deteriorate gradually beyond that timeframe. In our analysis and data presentation, we have only included spin data acquired within the first few hours after magnetization, where the remanent magnetization was well preserved. The spin-integrated ARPES spectra and the spin-resolved EDC were taken at 27 eV and $T = 6$ K. The Sherman function value for the mini Mott detectors is 0.12.

ASSOCIATED CONTENT

**Data Availability Statement:** All data are available in the main article and Supporting Information from the corresponding author upon reasonable request.

**Supporting Information:** Includes detailed growth characterization (section 1), band gap analysis techniques (low temperature data in section 2 and high temperature data in section 4), and high-temperature spin-ARPES spectra (section 3), along with accompanying Figures (PDF).

The Supporting Information is available free of charge at https://pubs.acs.org/....

AUTHOR INFORMATION

**Corresponding Author**

**Oliver J. Clark** - School of Physics and Astronomy, Monash University, Clayton, VIC 3800, Australia. Email: oliver.clark1@monash.edu

**Mark T. Edmonds** - School of Physics and Astronomy, Monash University, Clayton, VIC 3800, Australia; ARC Centre for Future Low Energy Electronics Technologies, Monash University, Clayton, VIC 3800, Australia. Email: mark.edmonds@monash.edu

**Author Contributions**

M.T.H.B., Q.L., and M.T.E. conceived the project. M.T.H.B., Q.L., and M.T.E. contributed to the materials design and growth. M.T.H.B., Q.L., O.J.C., and M.T.E. performed the characterization and measurements. J.E.L., J.D., J.S.B., A.F., E.R., S.K.M., and A.F. supported the beamline measurement facilities. O.J.C. and A.T. co-supervised the project. M.T.E. supervised the overall project and coordinated facility support. M.T.H.B. and O.J.C. wrote the manuscript. Q.L., J.B., and S.K.M. contributed to the manuscript writing. All authors approved the final version of the manuscript. †M.T.H.B. and Q.L. contributed equally to this work.

**Notes**

The authors declare no competing financial interest.

ACKNOWLEDGMENT

M. T. E. acknowledges funding support from ARC Future Fellowship FT2201000290. M.T.E., and M. T. H. B., acknowledge funding support from ARC Discovery Project DP250100026. O. J. C. acknowledges funding support from ARC Discovery Project DP200101345. M.T.E., O.J. C., Q. L., and M. T. H. B. acknowledge travel funding provided by the International Synchrotron Access Program (ISAP) managed by the Australian Synchrotron, part of ANSTO, and funded by the Australian Government. This research used resources of the Advanced Light Source, which is a DOE Office of Science User Facility under contract no. DE-AC02-05CH11231. M. T. H. B. and J. B. acknowledge funding support from the AINSE postgraduate award.

REFERENCES

(1) Hasan M. Z.; Kane C. L. Colloquium: Topological insulators. *Rev. Mod. Phys.* **2010**, *82*, 3045.

(2) Tokura Y.; Yasuda K.; Tsukazaki A. Magnetic topological insulators. *Nat. Rev. Phys.* **2019**, *1*, 126–143.

(3) Hsieh D.; Xia Y.; Qian D.; Wray L.; Dil J.; Meier F.; Osterwalder J.; Patthey L.; Checkelsky J.; Ong N.; Fedorov A.; Lin H.; Bansil A.; Grauer D.; Hor Y.; Cava R.; Hasan M. A tunable topological insulator in the spin helical Dirac transport regime. *Nature* **2009**, *460*, 1101–1105.

(4) Chang C. Z.; Zhang J.; Feng X.; Shen J.; Zhang Z.; Guo M.; Li K.; Ou Y.; Wei P.; Wang L. L.; Ji Z. Q.; Feng Y.; Ji S.; Chen X.; Jia J.; Dai X.; Fang Z.; Zhang S. C.; He Y. W. L. K.; Ma X. C.; Xue Q. K. Experimental observation of the quantum anomalous Hall effect in a magnetic topological Insulator. *Science* **2013**, *340*, 6129, 167-170.

(5) Chen Y.; Xu L.; Li J.; Li Y.; Wang H.; Zhang C.; Li H.; Wu Y.; Liang A. Topological Electronic Structure and Its Temperature Evolution in Antiferromagnetic Topological Insulator MnBi2Te4. *Phys. Rev.* **2019**, *X9*, 041040.

(6) Xia Y.; Qian D.; Hsieh D.; Wray L.; Pal A.; Lin H.; Bansil A.; Grauer D.; Hor Y. S.; Cava R. J.; Hasan M. Z. Observation of a large-gap topological-insulator class with a single Dirac cone on the surface. *Nat. Phys.* **2009**, *5*, 398–402.

(7) Yu R.; Zhang W.; Zhang H. J.; Zhang S. C.; Dai X.; Fang Z. Quantized anomalous Hall effect in magnetic topological insulators. *Science* **2010**, *329*, 5987, 61–64.

(8) Katmis F.; Lauter V.; Nogueira F. A high-temperature ferromagnetic topological insulating phase by proximity coupling. *Nature* **2016**, *533*, 7604, 513–516.

(9) Liu C. X.; Qi X. L.; Dai X.; Fang Z.; Zhang S. C. Quantum Anomalous Hall Effect in $Hg_{1-y}Mn_yTe$ Quantum Wells. *Phys. Rev. Lett.* **2008**, *101*, 146802.

(10) Qi X. L.; Zhang S. C. Topological insulators and superconductors. *Rev. Mod. Phys.* **2011**, *83*, 1057.

(11) Mong R. S. K.; Essin A. M.; Moore J. E. Antiferromagnetic topological insulators. *Phys. Rev. B* **2010**, *81*, 245209.

(12) Li Q.; Trang C. X.; Wu W.; Hwang J.; Cortie D.; Medhekar N.; Mo S. K.; Yang S. A.; Edmonds M. T. Large Magnetic Gap in a Designer Ferromagnet–Topological Insulator–Ferromagnet Heterostructure. *Adv. Mater.* **2022**, *34*, 2107520.

(13) Yao X.; Cui Q.; Huang Z.; Yuan X.; Yi H. T.; Jain D.; Kisslinger K.; Han M. G.; Wu W.; Yang H.; Oh S. Atomic-Layer-Controlled Magnetic Orders in $MnBi_2Te_4$–$Bi_2Te_3$ Topological Heterostructures. *Nano Lett.* **2024**, *24*(32), 9923−9930.

(14) Lee S. H.; Zhu Y.; Wang Y.; Miao L.; Pillsbury T. Spin scattering and noncollinear spin structure-induced intrinsic anomalous Hall effect in antiferromagnetic topological insulator $MnBi_2Te_4$. *Phys. Rev. Res.* **2019**, *1*, 012011.

(15) Sánchez-Barriga J.; Varykhalov A.; Springholz G.; Steiner H.; Kirchschlager R. Nonmagnetic band gap at the Dirac point of the magnetic topological insulator $(Bi_{1-x}Mn_x)_2Se_3$. *Nat. Commun.* **2016**, *7*, 10559.

(16) Yilmaz T.; Pertsova A.; Hines W.; Vescovo E.; Kaznatcheev K.; Balatsky A. V.; Sinkovic B. Gap-like feature observed in the non-magnetic topological insulators. J. Phys.: *Condens. Matter.* **2020**, *32*, 14.

(17) Zhang P.; Richard P.; Qian T.; Xu Y. M.; Dai X.; Ding H. A precise method for visualizing dispersive features in image plots. *Rev. Sci. Instrum.* **2011**, *82*, 043712.

(18) Trang C. X.; Li Q.; Yin Y.; Hwang J.; Akhgar G.; Bernardo I. D.; Grubišić-Čabo A.; Tadich A.; Fuhrer M. S.; Mo S. K.; Medhekar N. V.; Edmonds M.T. Crossover from 2D Ferromagnetic Insulator to Wide Band Gap Quantum Anomalous Hall Insulator in Ultrathin $MnBi_2Te_4$. *ACS Nano* **2021**, *15*(8), 13444–13452.

(19) Pan Z. H.; Vescovo E.; Fedorov A. V.; Gardner D.; Lee Y. S.; Chu S.; Gu G. D.; Valla T. Electronic Structure of the Topological Insulator $Bi_2Se_3$ Using Angle-Resolved Photoemission Spectroscopy: Evidence for a Nearly Full Surface Spin Polarization. *Phys. Rev. Lett.* **2011**, *106*, 257004.

(20) Jozwiak C.; Chen Y. L.; Fedorov A. V.; Analytis J. G.; Rotundu C. R.; Schmid A. K.; Denlinger J. D.; Chuang Y. D.; Lee D. H. Widespread spin polarization effects in photoemission from topological insulators. *Phys. Rev. B* **2011**, *84*, 165113.

(21) Fu L. Hexagonal Warping Effects in the Surface States of the Topological Insulator $Bi_2Te_3$. *Phys. Rev. Lett.* **2009**, *103*, 266801.

(22) Souma S.; Kosaka K.; Sato T.; Komatsu M.; Takayama A.; Takahashi T.; Kriener M.; Segawa K.; Ando Y. Direct Measurement of the Out-of-Plane Spin Texture in the Dirac-Cone Surface State of a Topological Insulator. *Phys. Rev. Lett.* **2011**, *106*, 216803.

(23) Alpichshev Z.; Analytis J. G.; Chu J. H.; Fisher I. R.; Chen Y. L.; Shen Z. X.; Fang A.; Kapitulnik A. STM Imaging of Electronic Waves on the Surface of $Bi_2Te_3$: Topologically Protected Surface States and Hexagonal Warping Effects. *Phys. Rev. Lett.* **2010**, *104*, 016401.

(24) Xu S. Y.; Neupane M.; Liu C.; Zhang D.; Richardella A.; Wray L. A. Hedgehog Spin-texture and Berry's Phase tuning in a Magnetic Topological Insulator. *Nat. Phys.* **2012**, *8*, 616–622.

(25) Li Q.; Bernardo I. D.; Maniatis J.; McEwen D.; Dominguez-Celorrio A.; Bhuiyan M. T. H.; Zhao M.; Tadich A.; Watson L.; Lowe B.; Vu T.-H.-Y.; Trang C. X.; Hwang J.; Mo S. K.; Fuhrer M. S.; Edmonds M. T. Imaging the Breakdown and Restoration of Topological Protection in Magnetic Topological Insulator $MnBi_2Te_4$. *Adv. Mater.* **2024**, *36*, 2312004.

(26) Rienks E. D. L.; Wimmer S.; Sánchez-Barriga J.; Caha O.; Mandal P. S.; Springholz G. Large magnetic gap at the Dirac point in $Bi_2Te_3$/$MnBi_2Te_4$ heterostructures. *Nature* **2019**, *576*(7787) 423–428.

(27) Han X.; Qu J.; Tan H.; Tao Z.; Meyer N. M.; Kirchmann P. S.; Guo Y.; Yan B.; Shen Z. -X.; Sobota J. A. Reconstructing the wave function of magnetic topological insulators $MnBi_2Te_4$ and $MnBi_4Te_7$ using spin-resolved photoemission. *Phys. Rev. X* **2025**, 15, 031022.

(28) Shikin A. M.; Zaitsev N. L.; Tarasov A. V.; Makarova T. P.; Glazkova D. A.; Estyunin D. A.; Klimovskikh I.I. Electronic and spin structure of topological surface states in $MnBi_4Te_7$ and $MnBi_6Te_{10}$ and their modification by an applied electric field. *Jetp Lett.* **2022**, 116, 556–566.